\documentclass[pre,twocolumn,twoside,showpacs,floatfix]{revtex4}
\usepackage{dsfont} 
\usepackage{amsmath}
\usepackage{epsfig}
\input epsf
\epsfclipon
\usepackage{epstopdf}

\begin{document}
\title{Spanning trees of the World Trade Web: \\real-world data and the gravity model of trade}
\author{Patryk Skowron, Mariusz Karpiarz, Agata Fronczak, Piotr Fronczak}
\affiliation{Faculty of Physics, Warsaw University of Technology, Koszykowa 75, PL-00-662 Warsaw, Poland}
\date{\today}

\begin{abstract}
In this paper, we investigate the statistical features of the weighted international-trade network. By finding the maximum weight spanning trees for this network we make the extraction of the truly relevant connections forming the network's backbone. We discuss the role of large-sized countries (strongest economies) in the tree. Finally, we compare the topological properties of this backbone to the maximum weight spanning trees obtained from the gravity model of trade. We show that the model correctly reproduces the backbone of the real-world economy.
\end{abstract}  \maketitle

\section{Introduction}

Since the seminal papers of Barab\'asi and Albert \cite{Science1999, RMP2002}, in which the authors showed that many socio-technical and natural systems have a very non-trivial, complex network structure, a large number of contributions were addressed, in which the methodology of complex networks was also used to study economic and financial systems \cite{ScienceEcon2009}. Various issues related to data analysis and modelling of financial and economic networks were also discussed at subsequent Polish Symposia on Econo- and Sociophysics (called in Polish FENS) (e.g. see~\cite{appaGworek2010, appaMiskiewicz2012, appaMiskiewicz2013, appaSienkiewicz2013}). In this paper we deal with the issue of spanning trees of the World Trade Web (WTW), which was presented at the 7th FENS in Lublin.

In the most general form, WTW is defined as the network of world-trade relations, where countries are represented by nodes and directed weighted links connecting them represent money flows from one country to another. In the last years, many stylized facts about WTW were reported, which, in accordance with the stage of development of the science of networks, were correspondingly related to: binary representation of this network \cite{physaLi2003, preSerrano2003, prlGarlaschelli2014}, its weighted version \cite{econSerrano2007, jspBhattacharya2008, preFagiolo2009}, multinetwork (commodity-specific) properties \cite{preBarigozzi2010}, the inherent community structure \cite{epjbTzekina2008}, and even fractal properties \cite{arxivFronczak2014}. Abundance of analyses and revealed stylized facts became the basis for theoretical models of WTW \cite{prlGarlaschelli2014, preFronczak2012, appbFronczak2012, njpMastrandrea2014}. And although, at present, the literature on WTW is quite extensive, the problem of spanning trees for this network appears there rather marginally (the few examples in this regard are~\cite{mst1,mst2,mst3}). The aim of this contribution is to address the maximum weight spanning trees for WTW in a more systematic way than in the previous works on this topic.

The paper is organized as follows. In Sect.~\ref{sec2}, we present the real data sets and introduce basic concepts and definitions that are in use throughout this article. In Sect.~\ref{sec3}, we discuss results of real-data analysis. In Sect.~\ref{sec4}, spanning trees of real WTW are compared with the corresponding spanning trees for synthetic, fully connected networks, in which connection weights are calculated according to the gravity model of trade. We summarize our results in Sect.~\ref{sec5}.

\section{Data description and basic definitions}\label{sec2}

Results described in this paper are based on the trade data collected by Gleditsch \cite{dataWTW} that contain, for each world country in the period 1950-2000, the detailed list of bilateral import and export volumes. The data are employed to build a sequence of symmetric matrices, $\mathbf{W}(t)$, corresponding to snapshots of weighted trade networks in the consecutive years, $t=1950,1951,\dots 2000$. Each entry, $w_{ij}(t)$, in any single matrix ,$\mathbf{W}(t)$, represents the average trade volume between $i$ and $j$ in a given year $t$. To be precise, $w_{ij}(t)$ is calculated as follows:
\begin{equation}\label{eq1}
w_{ij}(t)=\frac{1}{4}\left(w_{ij}^e(t)+w_{ij}^i(t)+w_{ji}^e(t)+w_{ji}^i(t)\right),
\end{equation}
where $w_{ij}^e(t)$ refers to the volume of export from $i$ to $j$, and $w_{ij}^i(t)$ stands for the volume of import from $i$ to $j$. From the point of view of an external observer, these two values should be the same. However, due to differences in reporting procedures between countries, there are often small deviations between $w_{ij}^e(t)$ and $w_{ij}^i(t)$. The same applies to trade in the opposite direction, thus justifying Eq.~(\ref{eq1}).
(For a detailed discussion about symmetry issues between $w_{ij}^e$ and $w_{ji}^e$ see e.g.~\cite{econFagiolo2006}.)

In this paper, maximum weight spanning trees for WTW, which are characterized by a sequence of matrices, $\mathbf{W^*}(t)$, with entries, $w_{ij}^*(t)$, are obtained by using the Prim's algorithm to graphs with connection weights equal to $-w_{ij}(t)$.

\begin{figure*}[ht]
\begin{center}
{\includegraphics[width=12cm]{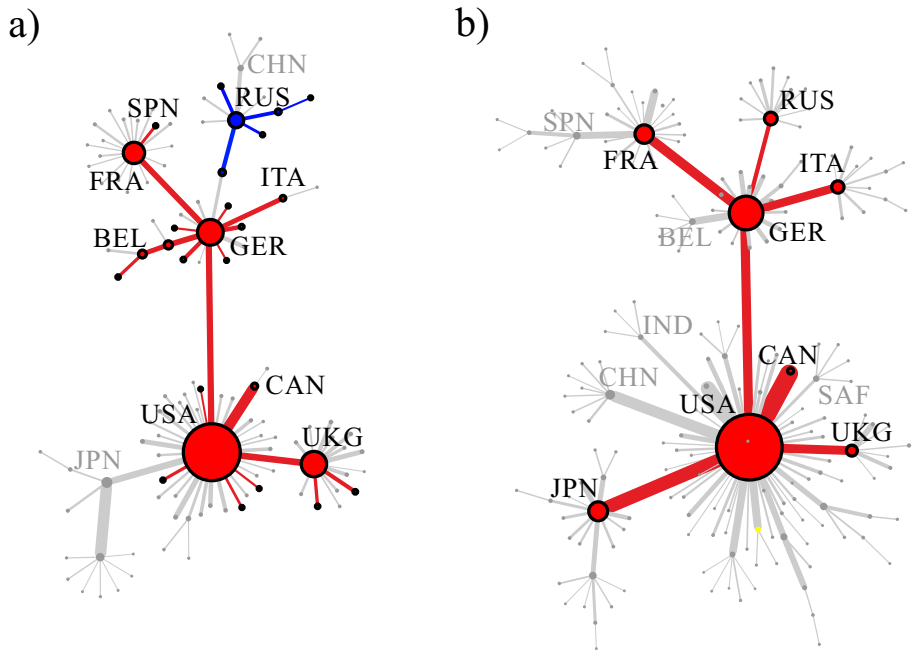}}
\caption{Maximum weight spanning trees for real WTW in: a) 1960 and b) 2000. Nodes in the trees are coloured according to the membership of the corresponding countries in various economic organizations (see description, which is given in the text).}\label{fig1mst}
\end{center}
\end{figure*}

Apart from trade matrices, we also use several other quantities that make description of structural properties of WTW easier. In particular, to characterize trade performance of a country we define the so-called strength, $s_i(t)$, of the corresponding node. The quantity is calculated as the total weight of all connections that are attached to the node and it represents the total export (or import) of the considered country:
\begin{equation}\label{eq2}
s_i(t)=\sum_{j=1}^{N(t)}w_{ij}(t),
\end{equation}
where $N(t)$ is the number of countries participating in the international trade in a given year. The total sum of the connections' weights in WTW is defined as:
\begin{equation}\label{eq3}
T(t)=\sum_{i=1}^{N(t)}\sum_{j=i+1}^{N(t)}w_{ij}(t)=\frac{1}{2}\sum_{i=1}^{N(t)}s_i(t),
\end{equation}
and the total number of such connections is given by:
\begin{equation}\label{eq4}
E(t)=\sum_{i=1}^{N(t)}\sum_{j=i+1}^{N(t)} \mathds{1}\!\left(w_{ij}(t)\right),
\end{equation}
where
\begin{equation}\label{deltawij}
\mathds{1}\!\left(w_{ij}(t)\right)=\begin{cases}1\;\;\;\mathrm{for}\;\;\;w_{ij}(t)>0
\\0\;\;\;\mathrm{for}\;\;\;w_{ij}(t)=0.\end{cases}
\end{equation} 
In the maximum weight spanning trees, the corresponding quantities: $s_i^*(t)$, $T^*(t)$, and $E^*(t)$ are defined in a similar manner, but using the entries of the matrix $\mathbf{W^*}(t)$ instead of $\mathbf{W}(t)$.

All the data used in this study are given in millions of current year U.S. dollars. The same applies to the trading countries' GDP (Gross Domestic Product) values, $\{x_i(t)\}$ \cite{dataGDP}. Finally, the distance between countries, $r_{ij}$, is the distance between their capitals, and it is given in kilometers \cite{dataR}.

\section{Real data analysis}\label{sec3}

\subsection{Graphical representation of WTW}

In Fig.~\ref{fig1mst}, maximum weight spanning trees for WTW in 1960 and 2000 are shown, which are obtained from real data. By analysing this figure one can understand, how geographical, political and historical conditions influence the global trade. In the spanning trees, the thickness of edges and the size of nodes reflect the corresponding bilateral trade volume, $w_{ij}$, and the strength of the country, $s_i$, respectively. The above means, that the bigger node is, the more significant is the role of the country in the international trade. In simple words, large nodes represent the stronger world economies. Such nodes (countries) usually have a higher number of nearest neighbours (star-like nodes), what distinguishes them from the less significant economies (leaf-like nodes with only one edge).

\begin{figure*}[ht]
\begin{center}
{\includegraphics[width=16cm]{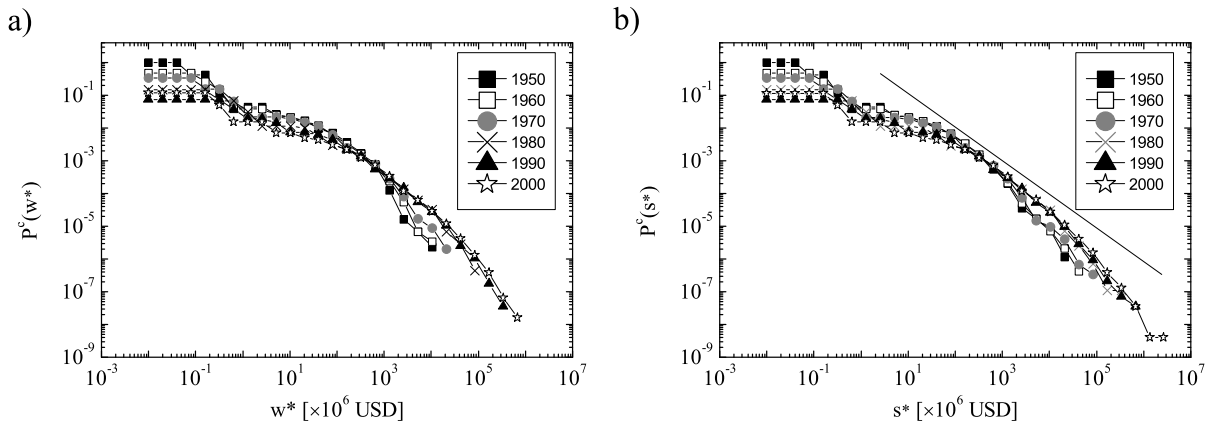}}
\caption{Weight and strength statistics of maximum weight spanning trees in different years. a) Cumulative distributions of trade volumes, $P^c(w^*)$, b) Cumulative distributions of nodes strength, $P^c(w^*)$.} \label{fig2}
\end{center}
\end{figure*}

It is clear from Fig.~\ref{fig1mst} that for the past 50 years the United States of America (USA) and Germany (GER, as the Federal Republic of Germany before 1990) have always played dominant roles in the international trade. At the same time, the economic importance of the other countries participating in trade changed (grown or decreased).

In particular, in 1960 (see Fig.~\ref{fig1mst}~a), among the most influential economies in the world were also: France (FRA, whose position was strong because of its colonies) and Russia (RUS, then as the Soviet Union, whose leading role was determined by the existence of the so-called Eastern Bloc along with a number of socialist states elsewhere in the world). In general, in the sixties, the structure of the international trade network was strongly dependent on the political zones of influence, which established after the Second World War and were reflected in the opposing economic organizations such as the Council for Mutual Economic Assistance (COMECON, 1949-1991, under the leadership of the Soviet Union, blue nodes) and the Organisation for European Economic Co-operation (OEEC, 1948-1961, bringing together the countries of Western Europe, which in 1961 was transformed into the Organisation for Economic Co-operation and Development, OECD, red nodes).

In a similar manner, when analysing the spanning tree of WTW in 2000 (see Fig.~\ref{fig1mst}~b), it becomes apparent that the backbone of WTW, which consists of star-like nodes, is created by the Group of Eight (G8) member countries, i.e. the United States (USA), Canada (CAN), Japan (JPN), Germany (GER), France (FRA), the United Kingdom (UKG), Italy (ITA), Russia (RUS) (red subtree in Fig.~\ref{fig1mst}~b). It was assumed that these countries represent the most developed economies in the world, with the largest GDP values and with the highest national wealths. It was not exactly the truth because G8 did not include China, which already were one of the fastest growing economy. Therefore, in order to increase the representativeness of the group, since 2005, there were meetings known as the G8+5, with representatives from China (CHN), Brazil, Mexico, India (IND), and South Africa (SAF). The importance of these countries in the world trade network topology is evident and is further discussed in subsequent sections of this paper.

\subsection{Weight and strength statistics}

WTW is a densely connected network \cite{preSerrano2003}. In Ref.~\cite{jspBhattacharya2008}, the authors argued that probability distribution of the weights of edges, $P(w)$, in this network is a log-normal distribution. They have shown that although the tail of the distribution is fat with significant fluctuations, when it is shown in a double logarithmic scale it reveals a clear curvature instead of a linear behaviour. Therefore, it can not be interpreted as a power-law. The similar findings were reported in subsequent studies, see e.g.~\cite{preFagiolo2009}.

In this section, we report on weight and strength statistics of the maximum weight spanning trees for WTW in the period 1950-2000. The cumulative versions of the corresponding distributions, $P^c(w^*)$ and $P^c(s^*)$, are shown in Fig.~\ref{fig2}. And although, the functional forms of the two distributions are questionable, it seems that the tail of the strength distribution can be described by a power-law, $P(s^*)\sim (s^*)^{-\gamma}$, with the time-independent characteristic exponent $\gamma\simeq 2$. The exponent is close to the exponent of the Pareto distribution, which is common to many other wealth distributions, including, for example, the distribution of GDP values of all countries in the world (see e.g. Fig.~1 in~\cite{prlGarlaschelli2014}).

For small values of $w^*$ and $s^*$ the considered distributions are almost identical. This is due to the fact that in this range, the two distributions describe leaves of the spanning tree, for which the following equality holds: $s_i^*=w_{ij}^*$, i.e. each leaf has only one trade channel. The less obvious conclusion drawn from the observed compatibility of distributions is that the strength of the nodes are positively correlated with weights of the attached edges (i.e. less developed economies have lower trade volumes).

\begin{figure*}[ht]
\begin{center}
{\includegraphics[width=16cm]{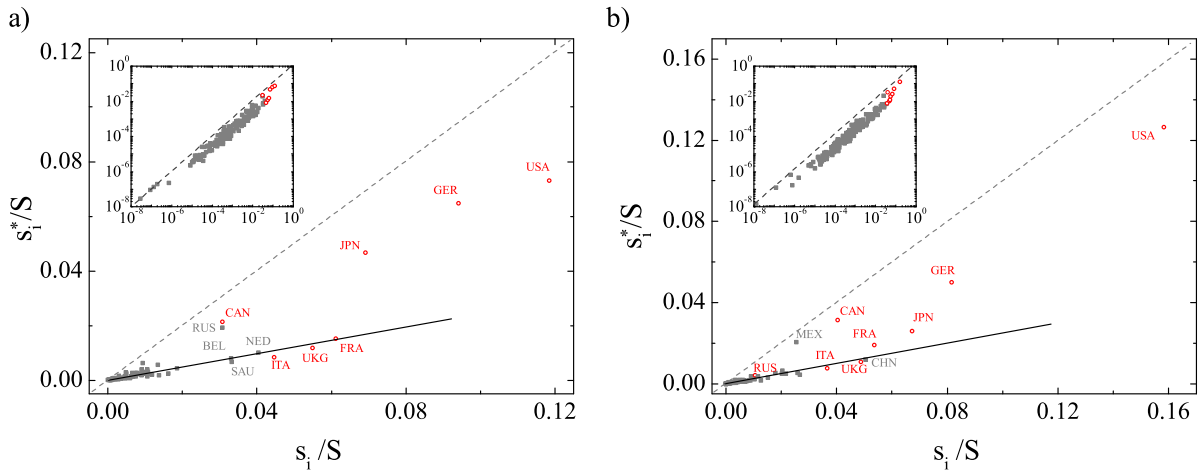}}
\caption{Relative strength, $s_i/S$, of nodes in WTW vs. their relative strength, $s_i^*/S$, in the maximum weight spanning tree in two different years: a) 1980,and b) 2000. Red points indicate the member states of the G7 (1980) and G8 (2000).}\label{fig3}
\end{center}
\end{figure*}

In Fig.~\ref{fig3}, the relative strength of each country, $s_i/S$ (i.e. $s_i$ divided by the strength of the whole network $S=\sum_is_i$), is shown in relation to the corresponding strength, $s_i^*/S$, in the spanning tree. The figure shows that nodes which represent different countries in the spanning tree can be roughly divided into two groups. The first group includes mainly those countries that are in the tree as leaves. They mostly have small GDP and, respectively, low trade performance (strength). The second and much less numerous group includes countries forming the skeleton of the tree. In the tree, they are usually represented by the star-like nodes. The first group, is characterized by the linear scaling relation: $s_i\sim s_i^*$. In the second group, the relation between $s_i$ and $s_i^*$ is not so obvious. However, if one wants to describe it as a linear relation, as in the first group, then the proportionality constant for the first group would be much smaller than for the second group. This indicates that, when creating the spanning tree, nodes of the first group lose more edges than nodes belonging to the second group. The reason may be that countries from different groups perform different functions in WTW. This makes the internal structure of WTW, which manifests itself in the maximum weight spanning tree, an interesting object to study.

\begin{figure}
{\includegraphics[width=8cm]{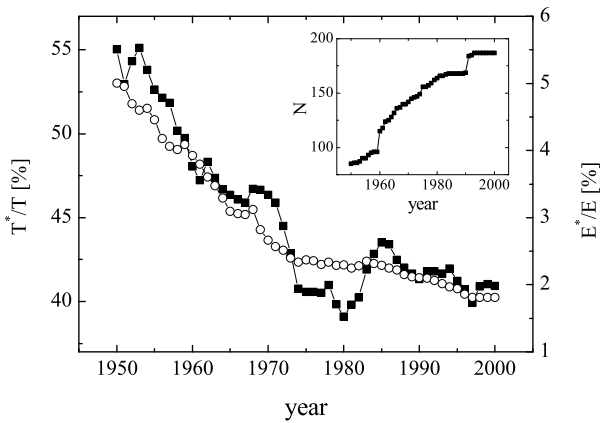}}
\caption{Time dependencies of: i) the percentage of the global trade volume which is covered by the tree $T^*/T$ (solid squares) and ii) the percentage of the number of all trade channels which are included in the tree $E^*/E$ (open circles). In the inset, the number of world countries vs. time is shown.}\label{fig4}
\end{figure}

\begin{figure}
{\includegraphics[width=7.05cm]{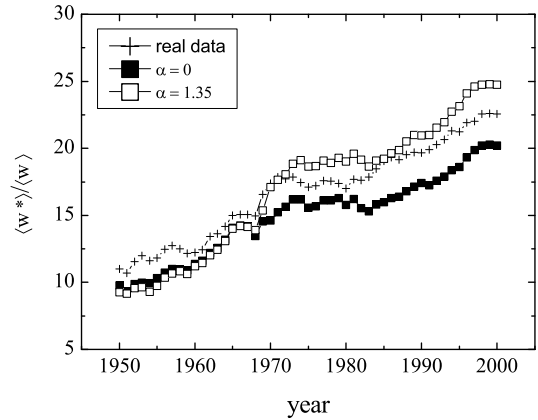}}
\caption{Time dependence of the ratio $\langle w^*\rangle/\langle w\rangle$, Eq.~(\ref{ratio}), between the average connection weight in the spanning tree, $\langle w^*\rangle$, and the corresponding average weight in the original WTW, $\langle w\rangle$. The figure shows data for real WTW and for two synthetic networks obtained from the gravity model of trade, Eq.~(\ref{gravity}), with two different values of the distance coefficient: $\alpha=0$ and $\alpha=1.35$.}\label{fig5}
\end{figure}

\subsection{Decreasing or increasing role of the tree}

\begin{figure*}[ht]
{\includegraphics[width=12cm]{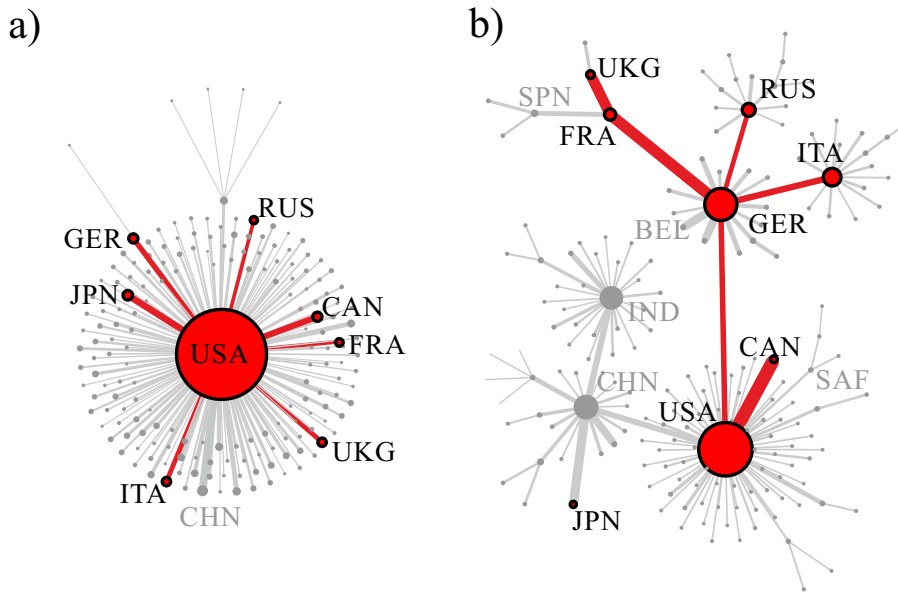}}
\caption{Maximum weight spanning trees for synthetic WTW obtained from the gravity model of trade, Eq.~(?), when assuming the same trade channels (the same binary network of trade connections) as in the real WTW in 2000. It is assumed that the distance coefficient in Eq.~(?) does not change over time and it is equal to: a) $\alpha=0$ and b) $\alpha=1.35$.}\label{fig6mst}
\end{figure*}

To examine, how the maximum weight spanning tree for the trade network has changed over time, we analysed time dependence of the following quantities: $T^*/T$ and $E^*/E$, which describe respectively: what percentage of the global trade volume is covered by the tree, and what percentage of the number of all trade channels is actually included in the tree. Fig.~\ref{fig4} shows that over the years 1950-2000, these parameters decrease monotonically. In particular, the total trade within the tree, $T^*$, which in the early fifties accounted for approximately $55\%$ of the global trade, $T$, in the late nineties accounted for only $40\%$ of $T$. Similarly, in the early fifties, the number of trade channels in the tree, $E^*$ was approximately $5\%$ of all trade channels, $E$, and in the late nineties, it was just $2\%$, thus indicating that in the analysed period of time many new trade connections emerged.

At first glance, Fig.~\ref{fig4} may indicate the declining role of the maximum weight spanning tree. On the other hand, however, when dividing $T^*/T$ by $E^*/E$ one gets the monotonically increasing ratio (see Fig.~\ref{fig5}): 
\begin{equation}\label{ratio}
\frac{\langle w^*\rangle}{\langle w\rangle}=\frac{T^*/E^*}{T/E}=\frac{T^*/T}{E^*/E},
\end{equation}
where $\langle w^*\rangle$ is the average connection weight in the spanning tree and $\langle w\rangle$ stands for the average weight in the original trade network. This, in turn, points to a completely different conclusion: Although the number of connections in WTW grows over time, these connections are not too significant. Over the past 50 years, the ratio $\langle w^*\rangle/\langle w\rangle$ grow linearly in time, indicating the increasing, not decreasing, role of the tree and proving that the tree can really be regarded as the backbone of WTW.

\section{Spanning trees obtained from the gravity model of trade}\label{sec4}

\subsection{Construction procedure for synthetic WTW}

To complete our study on maximum weight spanning trees for WTW, we have investigated whether the famous gravity model of trade, which is the basic macroeconomic model of the international trade, can be used to reproduce the spanning trees of real networks. 

The gravity model of trade was first proposed in 1962 by Jan Tinbergen, the physicist and the future first Nobel Prize Winner in Economic Sciences. Now, the model is one of the most recognizable empirical models in economics \cite{gravity1,gravity2,gravity3}. Drawing from Newton’s law of gravity, the gravity model relates the expected trade volume, $\langle w_{ij}(t)\rangle$, between two countries, $i$ and $j$, positively to the product of their GDP’s, i.e. $x_i(t)x_j(t)$, and negatively to the geographic distance, $r_{ij}$, between them. The simplest form of the gravity equation for the bilateral trade volume is:
\begin{equation}\label{gravity}
\langle w_{ij}(t)\rangle=G\;\frac{x_i(t)x_j(t)}{r_{ij}^{\alpha(t)}},
\end{equation}
where $G$ is a constant and $\alpha(t)$ is the distance coefficient, which is obtained from the real data analysis and which was recently identified as being the fractal dimension of the trade system \cite{arxivFronczak2014}.

As defined by Eq.~(\ref{gravity}), in the gravity model of trade, one assumes that each country has a trade connection with any other country (i.e. $\langle w_{ij}\rangle$ is non-zero for all pairs of countries). By this, synthetic trade networks (whose spanning trees we want to study), which would have been constructed under the gravity law of trade, would be fully connected graphs. This is quite unrealistic. Therefore, to make our study more reliable, we have decided to analyse only those trade channels, which are realized in real networks. More precisely: The studied synthetic networks have the same binary structure of trade connections as real WTW, but weights of these connections are calculated according to Eq.~(\ref{gravity}).

\begin{figure}
{\includegraphics[width=8cm]{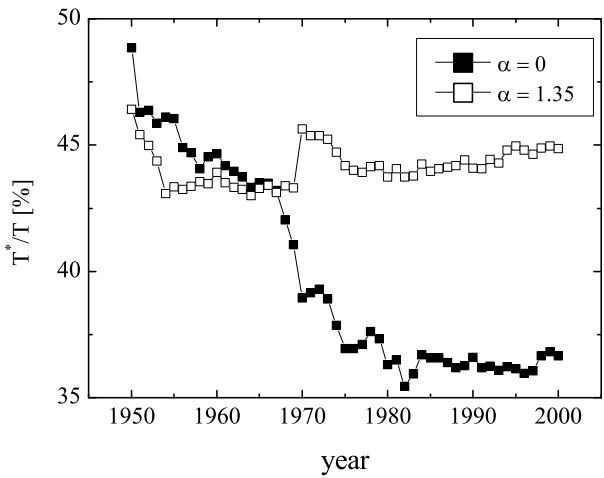}}
\caption{Time dependence of the percentage of the global trade
volume which is covered by the tree in synthetic trade networks.}\label{fig7}
\end{figure}

The above description means that, when studying gravity-based synthetic trade networks we employed the trading countries' GDP values, $\{x_i(t)\}$, and distances between their capitals, $r_{ij}$, to built a sequence of matrices $\mathbf{V}(t)$, whose entries were given by:
\begin{equation}\label{pompom}
v_{ij}(t)=G\;\frac{x_i(t)x_j(t)}{r_{ij}^\alpha}\; \mathds{1}\!\left(w_{ij}(t)\right),
\end{equation}
where $G=1$ was used, the distance coefficient $\alpha$ was assumed to be time-independent and equal to $0$ or $1.35$, and $\mathds{1}\!\left(w_{ij}(t)\right)$ was given by Eq.~(\ref{deltawij}). To justify the value of $G$, we would like to note that the precise value of this parameter is irrelevant in this study: $G$ does not affect structure of the tree, nor the ratios: $T^*/T$, $E^*/E$, and $\langle v^*\rangle/\langle v\rangle$. (Symbols for the global trade, $T$ and $T^*$, and the number of all connections, $E$ and $E^*$, in synthetic and real trade networks are the same.) When it comes to the distance coefficient, the values of $\alpha$ are meaningful in the sense that: $\alpha=0$ means no dependence on the distance, while $\alpha=1.35$ is the average value of this parameter in the period 1950-2000 \cite{arxivFronczak2014}. Finally, maximum weight spanning trees for synthetic WTW were obtained by using the Prim's algorithm to graphs with connection weights equal to $-v_{ij}$.

\subsection{Results of comparison between real and synthetic spanning trees}

In Fig.~\ref{fig6mst}, maximum weight spanning trees for synthetic gravity-like trade networks in 2000 are shown for two different values of the distance coefficient $\alpha=0$ and $\alpha=1.35$. The two trees shown are very different from each other. When $\alpha=0$, the tree has the form of a star, in which all countries are connected with the strongest economy in the world, i.e. the United States of America (USA). However, when the role of distance in trade is taken into account by using $\alpha=1.35$, then the spanning tree takes the form, which is very similar to the one which is shown in Fig.~\ref{fig1mst}~b. 

The remarkable difference between the two trees, Fig.~\ref{fig1mst}~b and Fig.~\ref{fig6mst}~b, is for connections between Asian countries. In the synthetic WTW, China (CHN) is seen as an economic power which dominates Asian trade network. In the real spanning tree, China's economic importance resulting from the reported trade volumes is much smaller. This difference may be due to the fact that the real network had no time to adapt to new external conditions: Trade is not keeping pace with the economic development of new economic powers such as China and India (IND).
This phenomenon can be compared to the phenomena of magnetic hysteresis, consisting in the fact that system's memory effects slow down the process of reaching the thermodynamic equilibrium.

In Fig.~\ref{fig7}, time dependence of the global trade volume which is covered by the synthetic tree, $T^*$, as compared with the total trade of the whole synthetic WTW, $T$, is shown for two values of $\alpha=0$ and $1.35$. Surprisingly, the results for $\alpha=0$ are much more similar to the corresponding results obtained for the real network (cf.~Fig.~\ref{fig4}). On the other hand, after dividing the obtained values of $T^*/T$ by $E^*/E$ (which are the same as in the real WTW), for both values of $\alpha$ one gets the ratios $\langle v^*\rangle/\langle v\rangle$ which are very similar to those obtained in the analysis of real networks.

\section{Summary}\label{sec5}

In this paper, we have studied some statistical features of the weighted international-trade network by means of maximum weight spanning trees. We have discussed the role of large-sized countries in the network's backbone explaining some geographical, political and historical conditions influencing the structure of this backbone. We have compared the topological properties of this backbone to the analogous one created from the gravity model of trade. We show that the model correctly reproduces the backbone of the real-world economics. We have suggested how memory effects in the trade system may influence the obtained results.

\section{Acknowledgements}
The work has been supported from the National Science Centre in Poland (grant no. 2012/05/E/ST2/02300).

\end{document}